# Explaining Negative Refraction without Negative Refractive Indices

GREGORY A. TALALAI,[1,*] TIMOTHY J. GARNER,[1] STEVEN J. WEISS[2]

[1]Department of Electrical Engineering, The Pennsylvania State University, University Park, Pennsylvania 16802, USA
[2]Army Research Laboratory, Adelphi, Maryland 20783, USA
*Corresponding author: gat131@psu.edu

*Negative refraction through a triangular prism may be explained without assigning a negative refractive index to the prism by using array theory. For the case of a beam incident upon the wedge, the array theory accurately predicts the beam transmission angle through the prism and provides an estimate of the frequency interval at which negative refraction occurs. Negative refraction effects not only occur in prisms made of traditional metamaterial unit cells, such as magneto-dielectric spheres, but also in solid prisms made of completely filled, homogenous unit cells. In both prisms, the hypotenuse has a staircase shape because they are built of cubic unit cells. The large phase delay imparted by each unit cell, combined with the staircase shape of the hypotenuse, creates the necessary conditions for negative refraction. Full wave simulations using the finite-difference time-domain method show that array theory accurately predicts the beam transmission angle for both solid prisms and prisms made of magneto-dielectric spheres.*

## 1. INTRODUCTION

Negative refraction from periodic structures has been observed in both physical experiments [1-3] and numerical computations [4,5]. Early on, it was shown that a composite medium of wires and split metal loops would present an effective negative index of refraction to waves of the appropriate polarization [6]. Shelby, Smith, and Schultz subsequently performed the original experiment demonstrating negative refraction in 2000 [1], when they measured the refracted beam from a triangular prism constructed of unit cells arranged into a staircased prism. In this case, if our purpose is only to reproduce the correct angle of transmission, the staircased prism may be modeled as a smooth, homogeneous prism made of a negative index material. Various attempts have been made to attach a deeper physical meaning to the negative index of refraction thus assigned. Many researchers, for example, compute effective medium properties from the induced average polarization and magnetization responses of the unit cells [7-9]. In particular, Liu and Alu have carried out calculations for the periodic medium composed of densely spaced magneto-dielectric spheres [4]. They conclude from their analysis that a dipolar-effective medium treatment predicts negative indices of refraction that are consistent with full-wave electromagnetic computations.

In contrast to these approaches, we develop a predictive model for negative refraction in triangular prisms based on finite array theory that does not depend on the prior determination of an appropriate effective medium description. Our model emphasizes the importance of a single property of the underlying unit cell of the periodic medium: the excess phase delay imparted by the unit cell on a wave during transmission across the cell, compared to the phase delay across an equal distance of free space. Starting from a knowledge of the excess phase delay and the geometry of the prism, we develop an equivalent linear array of aperture radiators. The main transmitted beam of the equivalent aperture array is equivalent to the main beam transmitted through the prism.

Using our model, we first revisit the staircased prism made of periodically arranged spherical magneto-dielectric unit cells as proposed by Liu and Alu [4]. We compare theoretical predictions of the beam transmission angle through the prism to those simulated in full-wave electromagnetic software. The agreement is excellent over the range of frequencies where the prism allows for beam transmission without appreciable reflections. The success of this model in predicting the occurrence of negative refraction suggests that, in fact, one need only select a unit cell with the correct phase delay to produce negative refraction. Thus, the model predicts that even a homogeneous prism, that is, a prism made of solid magneto-dielectric, will be negatively refracting provided we achieve the necessary phase delay through each sub-block forming the steps of the prism staircase. Results of full-wave simulations confirm this hypothesis.

Lastly, we point out some limitations inherent in a simple effective medium description of the triangular prism, and indicate how these limitations are overcome by the array-theoretic model. We note that the application of effective medium theory to describe the homogeneous prism is still possible, but that the boundaries of the unit cell must be selected with care. Even then, the effective medium model is not as accurate as the array-theoretic model.

## 2. THEORY

The behavior of a negatively refracting, triangular metamaterial prism can be explained by combining antenna array theory with a suitable field equivalence theorem. We consider a triangular prism with interior wedge angle $\alpha$ constructed with cubic unit cells with edge length $d$. The prism is not a perfect triangle, as its hypotenuse is approximated as a staircase that proceeds in steps according to the unit cell edge lengths. If we restrict ourselves to steps one cell in height, and $R$ cells wide, then we can achieve interior wedge angles

$$\alpha = \cot^{-1} R = 45°, 26.6°, 18°, \ldots; \quad R = 1, 2, 3, \ldots \qquad (1)$$

In our examples, we set R=4, giving an interior wedge angle $\alpha \approx 14°$.

We presume an electromagnetic plane wave is incident upon the long, flat edge of the triangular prism, pictured in Figure 1, and we measure the beam transmission angle $\phi$ counterclockwise from the positive $\hat{x}$ axis. The prism is negatively refracting if $\phi > \alpha + \pi/2$. We define a refraction angle $\theta = \frac{\pi}{2} + \alpha - \phi$, measured from the normal line to the hypotenuse. Negative refraction is characterized by beam transmission at angles $\theta < 0$.

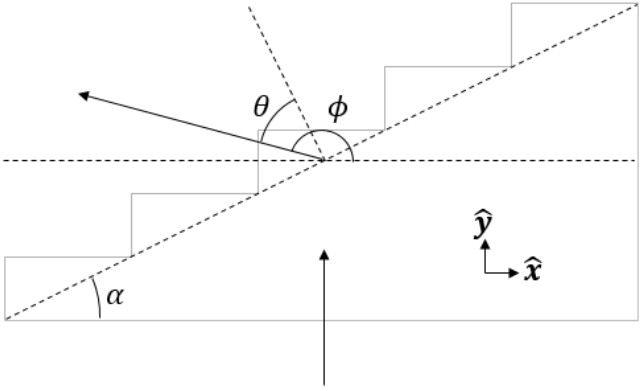

**Fig. 1**: Prism geometry.

Next, we draw the surface $S$ over the prism and divide this surface into N strips of equal width, according to Figure 2.

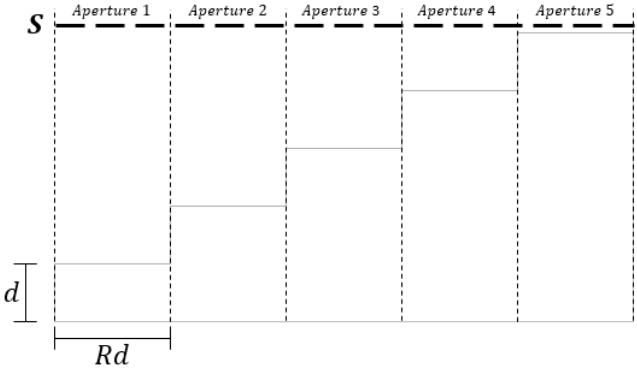

**Fig. 2**: Array theory geometry

We imagine each column as a transmission line feeding an aperture radiator situated upon its terminal edge. By invoking a field equivalence principle [10], we may replace the wedge with equivalent magnetic surface currents lying on $S$ radiating into free space. We assume that the currents on all terminal edges are identical except for their phases. This allows us to model the metamaterial prism as a scanning antenna array and to compute the beam transmission angle using the element pattern and array factor.

In our formulation, the key parameter of the metamaterial unit cell is the excess phase delay it delivers at each frequency, compared to an equal distance of free space. For instance, suppose the unit cell was a solid block of magneto-dielectric material with relative permeability $\mu_r$ and relative permittivity $\varepsilon_r$. The excess phase delay of the unit cell in this case would be

$$\psi = (k - k_0)d \qquad (2)$$

where $k = \sqrt{\mu_r \varepsilon_r} k_0$ and $k_0 = 2\pi/\lambda_0$ are the magneto-dielectric and free space wavenumbers, respectively, $\lambda_0$ is the free space wavelength, and $d$ is the length of any side of a single cubic unit cell. For densely packed magneto-dielectric spheres or other nontrivial unit cells, $\psi$ is determined from numerical simulations of a unit cell with suitably chosen periodic boundary conditions.

The number of unit cells encountered along each column is one more than the previous column. Thus, the transmission lines are configured to deliver a progressive phase delay to the radiating elements in the antenna array. We suppose that the transmitted electric field through the prism, along the surface $S$, is

$$\boldsymbol{E} = \hat{\boldsymbol{z}} E(x') \qquad (3)$$

where

$$E(x') = E_0 e^{-j(n-1)\psi} \quad ; (n-1)Rd < x' < nRd \qquad (4)$$

with $n = 1,2,\ldots,N$. Here, we have approximated the electric field on each aperture as uniform.

Invoking the perfect electric conductor (PEC) exterior equivalence [10], we place magnetic surface currents on $S$ equal to:

$$\boldsymbol{M}_{s,pec} = -\hat{\boldsymbol{y}} \times \boldsymbol{E} = -\hat{\boldsymbol{x}} E(x') \qquad (5)$$

which radiate in the presence of a PEC surface lying on $S$. If we apply image theory (ignoring the finite dimension of the wedge), we may image the magnetic surface currents over $S$, and obtain the currents

$$\boldsymbol{M}_s = -2\hat{\boldsymbol{y}} \times \boldsymbol{E} = -\hat{\boldsymbol{x}} 2E(x') \qquad (6)$$

which radiate into 2-dimensional free space, with the PEC surface removed. The far-field electric vector potential arising from these equivalent currents may be obtained by integrating over $S$ with the asymptotic form of the 2D Green's function. The 2D Green's function is given by

$$g(\boldsymbol{r}, \boldsymbol{r}') = \frac{1}{4j} H_0^{(2)}(k_0 |\boldsymbol{r} - \boldsymbol{r}'|)$$
$$\sim \frac{1}{4j} \sqrt{\frac{2j}{\pi k_0 \rho}} e^{-jk_0 \rho} e^{jk_0 x' \cos \phi} \qquad (7)$$

where $H_0^{(2)}(\ )$ is the Hankel function of the second kind and order 0, and $\rho$ is the radial distance in a cylindrical coordinate system. We have:

$$\boldsymbol{F} = \varepsilon_0 \int_0^{NRd} \boldsymbol{M}_s(x') g(\boldsymbol{r}, \boldsymbol{r}') dx' \qquad (8)$$

Using the asymptotic form of the Green's function, we obtain

$$\boldsymbol{F} = \varepsilon_0 \frac{1}{\sqrt{8\pi j k_0}} \frac{e^{-jk_0 \rho}}{\sqrt{\rho}} \int_0^{NRd} \boldsymbol{M}_s(x') e^{jk_0 x' \cos \phi} dx' \qquad (9)$$

We substitute (6) into (9), obtaining

$$F = -\hat{x}\varepsilon_0 \frac{E_0}{\sqrt{2\pi jk_0}} \frac{e^{-jk_0\rho}}{\sqrt{\rho}} \times$$

$$\sum_{n=1}^{N}\left[e^{-j(n-1)\psi}e^{j(n-1)Rk_0d\cos\phi}\int_0^{Rd}e^{jk_0x'\cos\phi}dx'\right] \quad (10)$$

Evaluating the remaining integral yields

$$F = -\hat{x}\varepsilon_0 \frac{RdE_0}{\sqrt{2\pi jk_0}} \frac{e^{-jk_0\rho}}{\sqrt{\rho}} \operatorname{sinc}\left(\frac{Rk_0d}{2}\cos\phi\right) \times$$

$$\sum_{n=1}^{N}\left[e^{-j(n-1)\psi}e^{j(n-1)Rk_0d\cos\phi}\right] \quad (11)$$

The far-field radiated electric field is given in terms of the electric vector potential by p. 284 [11]:

$$E = \frac{jk_0}{\varepsilon_0}\hat{\rho} \times F \quad (12)$$

where $\hat{\rho}$ is the radial unit vector in the cylindrical coordinate system. Substituting the right side of (11) into (12) yields

$$E = \hat{z}\, P(\phi) \cdot AF(\phi) \cdot \frac{e^{-jk_0\rho}}{\sqrt{\rho}} \quad (13)$$

where $P(\phi)$ is the element pattern resulting from uniform currents on the terminal edges

$$P(\phi) = \frac{RdE_0}{\sqrt{2\pi jk_0}}\operatorname{sinc}\left(\frac{Rk_0d}{2}\cos\phi\right)\sin\phi \quad (14)$$

and $AF(\phi)$ is the array factor

$$AF(\phi) = \sum_{n=1}^{N} e^{-j(n-1)\psi}e^{j(n-1)Rk_0d\cos\phi} \quad (15)$$

The array factor is maximized when

$$Rk_0d\cos\phi - \psi = 2m\pi\,;\ \ m = 0,\pm 1,\pm 2,\ldots \quad (16)$$

The transmitted beam pattern, according to (13-15) is a function of both the element pattern and the array factor. While the effect of the element pattern needs to be incorporated for an accurate determination of the actual angle of refraction, its contribution is secondary in importance to that of the array factor. Typically, array theory interprets (16) solved for $m = 0$ as a main beam while solutions for $m = \pm 1, \pm 2, \ldots$ are interpreted as grating lobes. For progressive phase delays less than $\pi$, grating lobe solutions are dependent on element spacing exceeding $\lambda_0/2$. However, with $\psi$ exceeding $\pi$, and spacing less than $\lambda_0/2$, it is entirely possible to obtain solutions for $m = \pm 1, \pm 2, \ldots$ - even when the $m = 0$ case cannot be solved. Such solutions may exist in angular directions described as negatively refracted beams.

For an approximate estimation of the frequency interval for the occurrence of negative refraction, we can temporarily disregard the effect of the element pattern. Accordingly, under this approximation, from (16), for negative refraction, $\phi > \alpha + \frac{\pi}{2} > \frac{\pi}{2}$; therefore $\cos(\phi) < 0$, and the first occurrence of negative refraction will be when $m = -1$. Setting $m = -1$, and solving (16) for $\phi$, we obtain

$$\phi = \cos^{-1}\left(\frac{\psi - 2\pi}{Rk_0d}\right) \quad (17)$$

Substituting (17) into the inequality $\alpha + \frac{\pi}{2} < \phi < \pi$ and solving the inequality for $\psi$ yields

$$2\pi(1 - d/\lambda_0 \cot\alpha) < \psi < 2\pi(1 - d/\lambda_0 \cos\alpha) \quad (18)$$

In the above inequality, both $\lambda_0$ and $\psi$ are functions of frequency. The inequality (18) can be utilized to estimate the frequency interval for negative refraction before engaging in detailed calculations. The estimate for the actual angle of refraction is obtained as the angle $\phi$ where the main beam within the transmitted pattern reaches a maximum, as calculated using (13-15).

## 3. NUMERICAL RESULTS

We performed numerical analysis of a 14° prism of magneto-dielectric spheres to test the theory proposed in Section II. The unit cell was a cube with 3.5 mm edges with a 1.57 mm-radius sphere in the center. The unit cell and prism design are based on those proposed by Liu and Alu [4]. The geometry of the unit cell is shown in Figure 3A, and the prism is shown in Figure 3B. The sphere was made of a hypothetical material with an index of refraction $n = \sqrt{\varepsilon_r\mu_r} = 12.3$ as proposed in [4], and the region of the cell outside of the sphere was free space. We examine the prism over the frequency range $0.514 < k_0d < 0.807$.

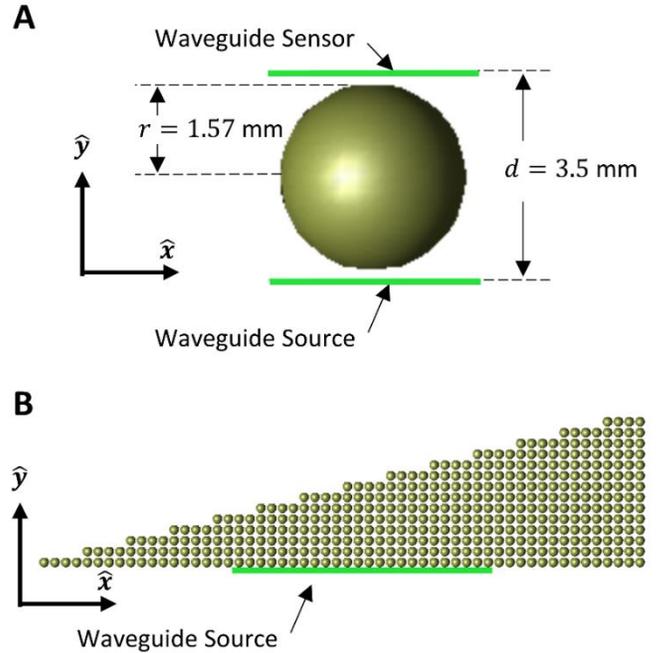

**Fig. 3**: Geometry of unit cell and prism. (A) Unit cell of 1.57 mm radius magneto-dielectric sphere with waveguide source and sensor used to measure the unit cell response. (B) 14° prism of magneto-dielectric spheres with waveguide source for simulation. Both panels show the coordinate axes of their respective geometries.

We use XFdtd [12], a commercial implementation of the finite-difference time-domain method [13,14], to perform our numerical experiments. The time-domain approach allows us to examine how negative refraction develops with time in addition to determining the steady-state refraction angles. The mesh grid is adaptive, with a mesh size of 0.16 mm inside and around the spheres and a coarser mesh grid with edge lengths of 2.0 mm in the regions away from the prism.

### A. Unit-Cell Response of Magneto-Dielectric Sphere

We compute the response of the unit cell by simulating it in a transverse electromagnetic (TEM) waveguide with perfect magnetic conductor (PMC) boundaries on the $\pm x$ faces and perfect electric conductor boundaries on the $\pm z$ faces. We placed TEM waveguide ports on the $\pm y$ faces of the unit cell, as shown in Figure 3A. The port on the $-y$ face transmits a Gaussian pulse, while the port on the $+y$ face serves as a sensor to measure the time-domain received signal. XFdtd uses the discrete Fourier transform to convert time-domain signal data to frequency-domain scattering parameters. We compute the excess phase delay of the unit cell by taking the difference between the unwrapped phase of the S21 scattering parameter and the phase delay of a cell filled with free space of the same size. The excess phase delay is shown in Figure 4. From (18) our wedge with $\alpha \approx 14°$ and unit cell sizes on the order of $d/\lambda_0 \approx 1/9$ will refract negatively when the excess phase delay lies approximately within the range $200° < \psi < 320°$. Using these numbers, Figure 4 predicts that negative refraction will occur within the frequency range of $0.679 < k_0 d < 0.761$.

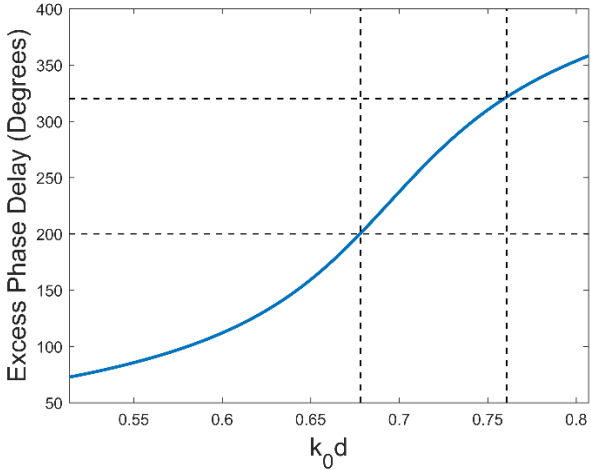

**Fig. 4**: Excess phase delay of unit cell with magneto-dielectric sphere vs. frequency, and expected negative refraction region (demarcated by dotted lines).

### B. Prism of Magneto-Dielectric Spheres

We model the 14° prism as a single layer of spheres with periodic boundary conditions on the $\pm z$ boundaries of the simulation to create the effect of an infinitely long prism. We use perfectly-matched layer absorbing boundary conditions on the other four sides of the simulation space. Our source is a TEM waveguide port polarized with its electric field aligned with the $y$ axis (and thereby the axis of the prism). The port is 24 unit cells wide and 1 unit cell high (the full height of the simulation space). We transmit a ramped-sinusoid waveform through the port and record the time-domain electric fields with a planar sensor on the $xy$ plane. Figure 3B shows the prism and waveguide ports along with the coordinate axes for the simulations.

Steady-state, instantaneous electric-field magnitudes are shown in Figure 5. Figure 5A shows the electric field for $k_0 d = 0.514$. At this frequency, the prism is positively refracting. Figures 5B and 5C show the negatively refracted beams at $k_0 d = 0.660$ and $k_0 d = 0.734$, respectively. At $k_0 d = 0.807$, the prism is again positively refracting, as is shown in Figure 5D.

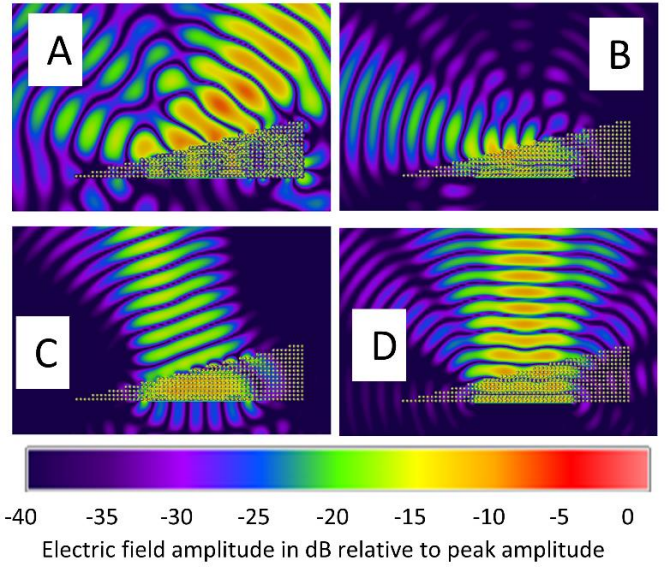

**Fig. 5**: Instantaneous electric fields refracted through 14° prism of magneto-dielectric spheres. All panels are scaled relative to the peak amplitude in that panel. (A) $k_0 d =$ 0.514 with peak amplitude of 5.1 kV/m. (B) $k_0 d =$ 0.660 with peak amplitude of 13.2 kV/m. (C) $k_0 d =$ 0.734 with peak amplitude of 10.5 kV/m. (D) $k_0 d =$ 0.807 with peak amplitude of 6.9 kV/m.

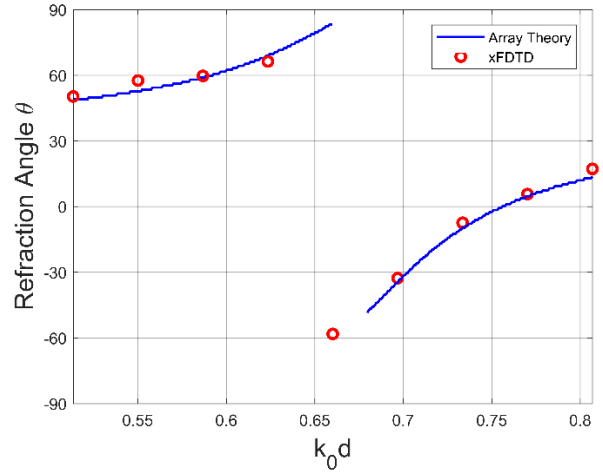

**Fig. 6**: Beam transmission angle through 14° prism of magneto-dielectric spheres: Array theory vs. xFDTD simulation.

Figure 6 compares the transmitted beam angle computed in XFdtd with the array-theoretic predictions. The results from XFdtd match the array-theoretic predictions within 5° over the range of $k_0 d$ studied. The XFdtd results show a transition from positive to negative refraction between $k_0 d = 0.624$ and 0.660, as predicted by the array theory model.

The theoretical calculation of the refracted angle is taken as the angle of peak transmitted electric field magnitude, as determined by the product of the element pattern (14) and the array factor (15). The progressive phase shift used in the array factor is shown in Figure 4. As the excess phase delay through each unit cell increases with frequency, $\theta$ increases until it reaches nearly 90°. At that point, $\theta$ is ill-defined as the beam is scanned through the nulls of the element pattern at $\phi = 0°$ and $\phi = 180°$. The array-theoretic model is somewhat unreliable over the range $0.660 < k_0 d < 0.679$, probably as a result of the rapid change of the beam with frequency. It is doubtful that the negatively refracting wedge could be physically realized over this frequency range, since the array factor is scanning close to the null directions of the element pattern.

The array-theoretic model accounts for interactions between cells in the prism by utilizing a phase delay for each unit cell calculated in the presence of a laterally infinite array of cells. Our theoretical model yields accurate estimates for the transmitted beam angle over a wide range of frequencies.

## C. Stepped-Hypotenuse Prism

According to our array-theoretic model, negative refraction may be achieved by appropriately selecting the excess phase delay. We are led to consider a cubic unit cell completely filled with magneto-dielectric material having positive real values for the relative permittivity and permeability. The macroscopic medium composed of solid cubic unit cells is itself a solid block of material that must be assigned a positive refractive index.

We simulate the solid prism with a stepped hypotenuse as shown in Figure 7A. The prism is made of the same material as the magneto-dielectric spheres in Section III B (i.e., of $n = 12.3$). Each cubic unit cell has dimension $d = 1.825$ mm. For this prism, $R = 4$, giving an interior wedge angle $\alpha \approx 14°$, identical to the prism with magneto-dielectric-sphere unit cells.

We simulate the stepped-hypotenuse prism in XFdtd at $k_0 d = 0.734$, where the phase delay in the material through the height of one step is 270°. We use a TEM waveguide interface similar to that used with the prism of spheres. Figure 7B shows the steady-state instantaneous electric field refracted through the prism. Negative refraction occurs with a beam transmission angle of $-52°$. The array theory predicts the occurrence of multiple beams, however one of these beams occurs at -51°, in agreement with the XFdtd results. Multiple beams appear in the array theory, which is consistent with the numerical simulation, and consistent with results shown in [5].

Since the prism is a solid block of positive index material, it is reasonable to inquire whether the effective medium theory has any merit. We could, for instance, conclude that the effective medium is the same magneto-dielectric material, but then predictions of the refraction angle will fail.

We may, however, apply the effective medium theory even to the solid negatively refracting wedge, as it offers some insight into the theory's applicability. Consider the unit cell to be a single cube of magneto-dielectric material of size $d$. In the presence of all neighboring "cells", we have an infinite slab of homogeneous magneto-dielectric. Thus, consider a plane wave propagating from the leading edge of this slab at $z = -d/2$ to the back edge at $z = +d/2$. This plane wave within the unit cell takes the form

$$\begin{aligned} \boldsymbol{E} &= \hat{\boldsymbol{x}} E_0 e^{-jkz} \\ \boldsymbol{H} &= \hat{\boldsymbol{y}} \frac{E_0}{\eta} e^{-jkz} \end{aligned} \quad (19)$$

The polarization and magnetization densities in the unit cell are

$$\begin{aligned} \boldsymbol{P} &= \epsilon_0 (\varepsilon_r - 1) \boldsymbol{E} = \hat{\boldsymbol{x}} \varepsilon_0 (\varepsilon_r - 1) E_0 e^{-jkz} \\ \boldsymbol{M} &= (\mu_r - 1) \boldsymbol{H} = \hat{\boldsymbol{y}} (\mu_r - 1) \frac{E_0}{\eta} e^{-jkz} \end{aligned} \quad (20)$$

The total electric and magnetic dipole moment resulting from the polarization and magnetization densities is obtained by integrating over the cubic volume of the unit cell. The densities have no variation in $x$ or $y$ so integration over these variables yields $d^2$, and all that remains is the integration over $z$.

$$\begin{aligned} \boldsymbol{p} &= \hat{\boldsymbol{x}} d^2 \varepsilon_0 (\varepsilon_r - 1) E_0 \int_{-d/2}^{+d/2} e^{-jkz} dz \\ \boldsymbol{m} &= \hat{\boldsymbol{y}} d^2 (\mu_r - 1) \frac{E_0}{\eta} \int_{-d/2}^{+d/2} e^{-jkz} dz \end{aligned} \quad (21)$$

Evaluating the remaining integrals yields

$$\begin{aligned} \boldsymbol{p} &= \hat{\boldsymbol{x}} d^3 \varepsilon_0 (\varepsilon_r - 1) E_0 \frac{\sin\left(\frac{kd}{2}\right)}{\frac{kd}{2}} \\ \boldsymbol{m} &= \hat{\boldsymbol{y}} d^3 (\mu_r - 1) \frac{E_0}{\eta} \frac{\sin\left(\frac{kd}{2}\right)}{\frac{kd}{2}} \end{aligned} \quad (22)$$

The effective medium parameters may then be obtained by taking the ratio of the average polarization and magnetization densities with the electric and magnetic field. The average polarization is $p/d^3$ and the average magnetization is $m/d^3$, where $p$ and $m$ are the amplitudes of the electric and magnetic dipoles given in (22). We obtain

$$\begin{aligned} \varepsilon_{r,\text{eff}} &= 1 + \frac{p}{\varepsilon_0 E_0 d^3} = 1 + (\varepsilon_r - 1) \frac{\sin\left(\frac{kd}{2}\right)}{\frac{kd}{2}} \\ \mu_{r,\text{eff}} &= 1 + \frac{m\eta}{E_0 d^3} = 1 + (\mu_r - 1) \frac{\sin\left(\frac{kd}{2}\right)}{\frac{kd}{2}} \end{aligned} \quad (23)$$

Substitution of the relevant parameters into (23) yields the result $\varepsilon_{r,\text{eff}} = -1.2$ and $\mu_{r,\text{eff}} = -1.8$. Provided the correct sign is chosen for the square root [15], the effective index of refraction is then

$$n_{\text{eff}} = -\sqrt{\mu_{r,\text{eff}} \varepsilon_{r,\text{eff}}} = -1.47 \quad (24)$$

The value observed from the simulation, consistent with the transmission angle of $-52°$ is $n_{\text{eff}} = -3.25$. Thus the effective medium theory gives the correct sign for the index of refraction, but the accuracy is poor. We hold out the possibility that the accuracy of the effective medium theory could be improved by incorporating advanced techniques [8,9], but the resulting models would be complicated. Moreover, the array theory is accurate without excessive complication. An essential advantage of the array theory model is that it accounts for the lateral extent of the excitation of the wedge in an approximate fashion through the parameter $N$.

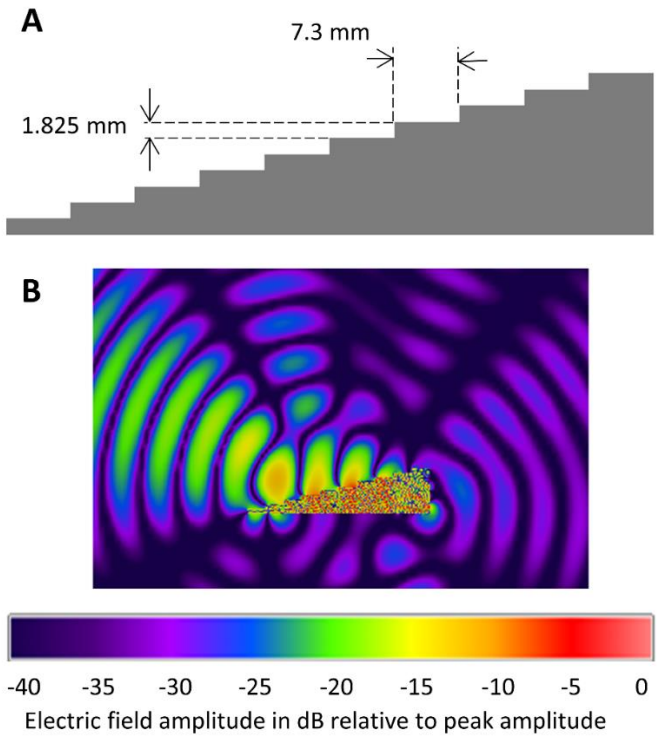

Figure 7: Solid prism with staircased hypotenuse. (A) Geometry of prism. (B) Instantaneous electric-field magnitude at 10 GHz after reaching steady state. Scale is relative to maximum electric-field amplitude of 59.4 kV/m.

## 4. CONCLUDING REMARKS

While both physical experiments [1-3] and numerical computations [4,5] have demonstrated the existence of negatively refracted beams, we have shown that negative refraction can be explained without assigning an effective index of refraction. In the case of the solid stepped-hypotenuse prism, in order to obtain appropriately negative constitutive parameters from effective medium theory, the sub-blocks of the prism must be treated as unit cells.

Array theory provides a simplified explanation of negative refraction and gives accurate predictions for all cases studied. Array-theoretic predictions of the frequency range for negative refraction, according to (17), were seen to depend on the interior wedge angle $\alpha$. The array-theoretic model accounts for dependencies of the refractive index of the prism upon the prism geometry itself. In particular, the model connects negative refraction with the staircasing of the prism, through the progressive phase shift in the equivalent aperture array. Furthermore, the array model takes into consideration the lateral extent of the impinging beam upon the wedge, and consequently predicts refraction angles with greater reliability than the effective medium theory.

The model also provides answers to objections raised by Valanju et al. [16] and Munk (Fig 1.11) [17]. Valanju et al. argue that negative refraction is impossible from negative index media using arguments based on causality and dispersion of such effective media. However, our model demonstrates negative refraction is not fundamentally related to an effective medium description. Moreover, the negatively refracted beams demonstrated here are formulated in the theory as far field phenomena associated with a linear array of radiators, and not as literal refraction. Provided we adopt this viewpoint, we need not explain any of the usual paradoxes arising from a consideration of the field behavior directly at the interface with the effective medium. Indeed, the in-phase addition needed to form the negatively refracted beam is simply understood to be an addition of wavefronts from each individual radiator offset by $2\pi$ (or integer multiples of $2\pi$). Accordingly, a number of cycles are needed before the formation of the beam [18].

Munk, on the other hand, does consider that negative refraction might result from the formation of a grating lobe in the far field, when a single sheet of unit cells is excited by a uniform plane wave. However, Munk rejects this interpretation because the lateral unit cell spacing for a valid effective medium description is at least smaller than a half-wavelength, precluding grating lobe formation. By contrast, in our model, each radiating aperture is preceded by a column of unit cells delivering a progressive phase shift independent of the lateral spacing between unit cells. This admits the possibility of a grating lobe, and by extension, of negatively refracted beams [18].

**Acknowledgment**. The authors thank Tarun Chawla and Walter Janusz from Remcom, Inc. (State College, PA, USA) for providing the XFdtd software and technical assistance with that software.

This research was conducted with Advanced CyberInfrastructure computational resources provided by the Institute for CyberScience at The Pennsylvania State University (http://ics.psu.edu).